\documentclass[twocolumn]{aastex631}

\usepackage{threeparttable}
\usepackage{amsmath} % Required for cases

\begin{document}

\title{An Ejection Event Captured by VLBI During the Outburst of Swift~J1727.8$-$1613}

\author[0000-0003-1514-881X]{Hongmin Cao}
\affiliation{School of Electronic and Electrical Engineering \\
Shangqiu Normal University, 298 Wenhua Road, Shangqiu, Henan 476000, PR China; \url{hongmin.cao@foxmail.com}}

\author[0000-0002-2322-5232]{Jun Yang}
\affiliation{Department of Space, Earth and Environment \\
Chalmers University of Technology, Onsala Space Observatory, 43992 Onsala, Sweden}

\author[0000-0003-3079-1889]{S{\'a}ndor Frey}
\affiliation{Konkoly Observatory, HUN-REN Research Centre for Astronomy and Earth Sciences\\
Konkoly Thege Mikl{\'o}s {\'u}t 15-17, 1121 Budapest, Hungary}
\affiliation{CSFK, MTA Centre of Excellence\\
Konkoly Thege Mikl{\'o}s {\'u}t 15-17, 1121 Budapest, Hungary}
\affiliation{Institute of Physics and Astronomy, ELTE E{\"o}tv{\"o}s Lor{\'a}nd University\\
P{\'a}zm{\'a}ny P{\'e}ter s{\'e}t{\'a}ny 1/A, 1117 Budapest, Hungary}

\author[0000-0002-2758-0864]{Callan M. Wood}
\affiliation{International Centre for Radio Astronomy Research, Curtin University, GPO Box U1987, Perth, WA 6845, Australia}

\author[0000-0003-3124-2814]{James C. A. Miller-Jones}
\affiliation{International Centre for Radio Astronomy Research, Curtin University, GPO Box U1987, Perth, WA 6845, Australia}

\author[0000-0003-1020-1597]{Krisztina {\'E}. Gab{\'a}nyi}
\affiliation{Department of Astronomy, Institute of Physics and Astronomy, ELTE E{\"o}tv{\"o}s Lor{\'a}nd University\\
P{\'a}zm{\'a}ny P{\'e}ter s{\'e}t{\'a}ny 1/A, 1117 Budapest, Hungary}
\affiliation{HUN-REN--ELTE Extragalactic Astrophysics Research Group, ELTE E{\"o}tv{\"o}s Lor{\'a}nd University\\
P{\'a}zm{\'a}ny P{\'e}ter s{\'e}t{\'a}ny 1/A, 1117 Budapest, Hungary}
\affiliation{Konkoly Observatory, HUN-REN Research Centre for Astronomy and Earth Sciences\\
Konkoly Thege Mikl{\'o}s {\'u}t 15-17, 1121 Budapest, Hungary}
\affiliation{CSFK, MTA Centre of Excellence\\
Konkoly Thege Mikl{\'o}s {\'u}t 15-17, 1121 Budapest, Hungary}
\affiliation{Institute of Astronomy, Faculty of Physics, Astronomy and Informatics\\
Nicolaus Copernicus University, Grudzi\c adzka  5, 87-100 Toru\'n, Poland}

\author[0000-0003-0216-8053]{Giulia Migliori}
\affiliation{INAF – Istituto di Radioastronomia, Via Gobetti 101, 40129 Bologna, Italy}

\author[0000-0002-8657-8852]{Marcello Giroletti}
\affiliation{INAF – Istituto di Radioastronomia, Via Gobetti 101, 40129 Bologna, Italy}

\author[0000-0003-0721-5509]{Lang Cui}
\affiliation{Xinjiang Astronomical Observatory, Chinese Academy of Sciences \\
150 Science 1-Street, Urumqi, Xinjiang 830011, PR China}

\author[0000-0003-4341-0029]{Tao An}
\affiliation{Shanghai Astronomical Observatory, Key Laboratory of Radio Astronomy \\
Chinese Academy of Sciences, 80 Nandan Road, Shanghai 200030, PR China}

\author[0000-0002-1992-5260]{Xiaoyu Hong}
\affiliation{Shanghai Astronomical Observatory, Key Laboratory of Radio Astronomy \\
Chinese Academy of Sciences, 80 Nandan Road, Shanghai 200030, PR China}

\author[0000-0002-9909-3758]{WeiHua Wang}
\affiliation{School of Computer Information Engineering, Changzhou Institute of Technology, Changzhou 213032, PR China}

%% Note that the \and command from previous versions of AASTeX is now
%% depreciated in this version as it is no longer necessary. AASTeX 
%% automatically takes care of all commas and "and"s between authors names.

%% AASTeX 6.31 has the new \collaboration and \nocollaboration commands to
%% provide the collaboration status of a group of authors. These commands 
%% can be used either before or after the list of corresponding authors. The
%% argument for \collaboration is the collaboration identifier. Authors are
%% encouraged to surround collaboration identifiers with ()s. The 
%% \nocollaboration command takes no argument and exists to indicate that
%% the nearby authors are not part of surrounding collaborations.

%% Mark off the abstract in the ``abstract'' environment. 
\begin{abstract}
We observed a newly-discovered Galactic black hole X-ray binary Swift~J1727.8$-$1613 with the European Very Long Baseline Interferometry Network (EVN) at 5~GHz. The observation was conducted immediately following a radio quenching event detected by the Karl G. Jansky Very Large Array (VLA). The visibility amplitude evolution over time reveals a large-amplitude radio flare and is consistent with an ejection event. The data can be interpreted either as a stationary component (i.e., the radio core) and a moving blob, or as two blobs moving away from the core symmetrically in opposite directions. The initial angular separation speed of the two components was estimated to 30~mas\,${\rm d}^{-1}$. We respectively fitted a single circular Gaussian model component to each of 14 sliced visibility datasets. For the case of including only European baselines, during the final hour of the EVN observation, the fitted sizes exhibited linear expansion, indicating that the measured sizes were dominated by the angular separation of the two components. The 6-h EVN observation took place in a rising phase of an even larger 4-day-long radio flare, implying that the ejection events were quite frequent and therefore continuous radio monitoring is necessary to correctly estimate the power of the transient jet. Combined with X-ray monitoring data, the radio quenching and subsequent flares/ejections were likely driven by instabilities in the inner hot accretion disk.

\end{abstract}

%% Keywords should appear after the \end{abstract} command. 
%% The AAS Journals now uses Unified Astronomy Thesaurus concepts:
%% https://astrothesaurus.org
%% You will be asked to selected these concepts during the submission process
%% but this old "keyword" functionality is maintained in case authors want
%% to include these concepts in their preprints.
%\keywords{stars:individual:Swift J1727.8--1623---ISM:jets and outflows---X-rays:binaries}
%\keywords{Classical Novae (251) --- Ultraviolet astronomy(1736) --- History of astronomy(1868) --- Interdisciplinary astronomy(804)}

%% From the front matter, we move on to the body of the paper.
%% Sections are demarcated by \section and \subsection, respectively.
%% Observe the use of the LaTeX \label
%% command after the \subsection to give a symbolic KEY to the
%% subsection for cross-referencing in a \ref command.
%% You can use LaTeX's \ref and \label commands to keep track of
%% cross-references to sections, equations, tables, and figures.
%% That way, if you change the order of any elements, LaTeX will
%% automatically renumber them.
%%
%% We recommend that authors also use the natbib \citep
%% and \citet commands to identify citations.  The citations are
%% tied to the reference list via symbolic KEYs. The KEY corresponds
%% to the KEY in the \bibitem in the reference list below. 

\section{Introduction}
\label{sec:intro}
  
Black hole low-mass X-ray binaries (BH LMXBs) are systems where the stellar-mass BH (i.e., the primary star) is accreting material from its companion, a main-sequence or evolved star with a mass less than $1\,{\rm M}_\sun$ \citep{Corral-Santana2016, Avakyan2023}. These systems spend most of their lives in quiescent state, and occasionally enter outburst phase due to thermal-viscous instabilities in the accretion disk. During an outburst, usually lasting for several months or years, BH LMXBs transition through different spectral states, signified by their X-ray spectral and timing characteristics \citep{Homan2005, Remillard2006}. These spectral states typically evolve from the low-hard state (LHS), through the hard intermediate state (HIS) and the short-lived soft intermediate state (SIS), to the high-soft state (HSS), and finally return to the LHS via the low-luminosity intermediate state \citep[e.g.][]{Motta2021}. The compact jet is always observed in the hard state, and the transient jet is produced when the source switches from HIS to SIS \citep{Fender2004, Fender2006, Fender2009, Bright2020, Carotenuto2021}. The radio astronomical technique of very long baseline interferometry (VLBI), offering the highest angular resolution, can investigate jet activity close to the BH and thus provide critical information for understanding the jet launching, acceleration, and collimation mechanisms \citep{Paragi2013, Yang2010, Yang2023, Miller-Jones2019, Wood2021, Wood2023}.

When the source is approaching the SIS from HIS, its radio emission generally decreases. The spectral state transition from HIS to SIS is characterized by a drop in fractional rms and a switch from type-C to type-B quasi-periodic oscillations (QPOs), as seen in the X-ray variability data \citep{Ingram2019}. Around this transition, large-amplitude radio flares are observed. Discrete ejecta, associated with the radio flare(s), are suspected to have a common physical origin with the type-B QPO. However, the inferred ejection times of the relativistic blobs can occur either before or after the emergence of the type-B QPO, suggesting that the two phenomena may not necessarily be linked to the same physical process \citep{Fender2009, Miller-Jones2012, Ingram2019, Belloni2016}. Radio flares, accompanied by ejection events, can also occur in the HIS, when the type-C QPO is always observed \citep[e.g.][]{Wood2025,Liao2024}.

Swift~J1727.8$-$1613 was discovered by the Burst Alert Telescope on board the Neil Gehrels Swift Observatory (Swift/BAT) on August 24, 2023 \citep{Page2023}. The rapid follow-up optical spectroscopy and radio observations classified this source as a Galactic BH LMXB candidate \citep{Negoro2023, Kennea2023, Nakajima2023, Castro-Tirado2023, Miller-Jones2023b}. The source reached the maximum flux of $6.75 \pm 0.06$~Crab in the $2-20$~keV band on August 29, 2023, and then entered the flux decay phase \citep{Matsuoka2009}. The sub-millimeter and optical polarization angles, measured in the first half of September 2023, were both close to zero, implying that the jet is aligned in the north--south direction \citep{Vrtilek2023, Kravtsov2023}. This was confirmed by VLBI imaging observations using the U.S. Very Long Baseline Array (VLBA) and the Australian Long Baseline Array (LBA) carried out in the same period \citep{Wood2024}. These observations detected a prominent jet extending to the south \citep{Wood2024}. The X-ray polarization degree of $3-4$\% and the polarization position angle oriented in the north--south direction, measured by the Imaging X-ray Polarimetry Explorer (IXPE) satellite in non-soft states, suggest a generally flat geometry of the hot disk oriented perpendicularly to the jet \citep{Veledina2023, Ingram2024, Podgorny2024}. In the soft state, the X-ray polarization degree showed a dramatic drop to $\lesssim 1$\% \citep{Svoboda2024}. The source completed the full state transition and entered the quiescent state in June 2024. The post-outburst optical spectroscopic observations constrained the binary orbital period of $\sim 10.8$~h, the mass lower limit of the primary star of $\sim 3.1\,{\rm M}_\sun$, and the source distance of $3.4 \pm 0.3$~kpc \citep{Mata2025}. This confirms the system being a BH LMXB. A new distance measurement is being carried out by \citet{Burridge2025}. At present, an accurate parallax distance is not available, and all the distance-related parameters in this Letter are calculated based on \citet{Mata2025}.

We proposed a single-epoch Target of Opportunity (ToO) observation with the European VLBI Network (EVN) in September 2023, to probe the jet properties of this extraordinarily bright X-ray source at milliarcsecond (mas) angular scales. These observations and the data reduction are described in Section~\ref{sec:obs}. The results are presented in Section~\ref{sec:resul}, and discussed in Section~\ref{sec:disc}. A summary is provided in Section~\ref{sec:sum}.

\section{Observations and Data Reduction} 
\label{sec:obs}

The EVN experiment was carried out in real-time electronic VLBI (eVLBI) mode \citep{Szomoru2004} at 5~GHz frequency using phase referencing \citep{Beasley1995}, starting from 12:30 UTC on October 5, 2023, and lasting for 6~h. This was right in between the radio flux density quenching to $\sim7$~mJy on October 5 and the dramatic rise to $\sim236$~mJy on October 6, observed by the Karl G. Jansky Very Large Array (VLA) at 5~GHz \citep{Miller-Jones2023a}. The dynamic power spectrum measured by the Neutron star Interior Composition Explorer (NICER) instrument on board the International Space Station showed a broad-band ($0.1-10$~Hz) fractional rms noise drop (from 10\% to 4\%), and the low-frequency quasi-periodic oscillation (QPO) became undetected during 3:00--11:00 UTC on October 5, suggesting a possible state transition from HIS to SIS \citep{Bollemeijer2023}.

\begin{figure}
    \centering
    \includegraphics[width=0.9\linewidth]{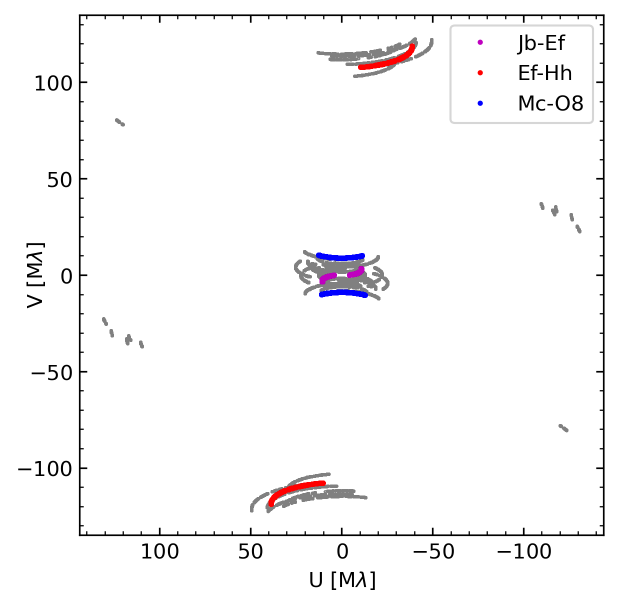}
    \caption{The $(u,v)$ coverage of the EVN experiment observing Swift~J1727.8$-$1613. The intra-European baselines are clustered at the center, while the longer baselines between the European antennas and Hh are roughly aligned along the north--south direction. The $(u,v)$ points belonging to the Jb--Ef, Mc--O8 and Ef--Hh baselines are highlighted by magenta, blue, and red colors, respectively.}
    \label{fig:1}
\end{figure}

The participating EVN antennas were Jodrell Bank Mk2 (Jb, United Kingdom), Westerbork (Wb, The Netherlands), Effelsberg (Ef, Germany), Medicina (Mc, Italy), Onsala (O8, Sweden), Tianma (T6, China), Toru\'{n} (Tr, Poland), Hartebeesthoek (Hh, South Africa), and Irbene (Ib, Latvia). The phase-reference cycle time was 7~min, including 4~min spent on the target. The bright quasar J1724$-$1443, at an angular separation of $1\fdg63$, was observed as phase-reference calibrator. The data were recorded at a rate of 2~Gbps, with left and right circular polarizations and eight 32-MHz subbands (intermediate frequency channels, IFs) per polarization, except for Wb where four 32-MHz subbands were used. The recorded data were transferred via optical fiber to the Joint Institute for VLBI European Research Infrastructure Consortium (JIVE), Dwingeloo, The Netherlands, and processed there with the SFXC software correlator \citep{Keimpema2015} with 64 spectral channels per IF and 2-s integration time. The $(u,v)$ coverage is shown in Fig.~\ref{fig:1}. 

%\section{Data reduction} 
%\label{sec:reduc}

The EVN data reduction guide\footnote{\url{https://www.evlbi.org/evn-data-reduction-guide}} was generally followed to reduce the data in the Astronomical Image Processing System \citep[AIPS,][]{Greisen2003}. The a priori amplitude calibration, parallactic angle correction, ionospheric correction, and instrumental delay correction were performed accordingly, followed by global fringe-fitting and bandpass calibration. The phase-reference calibrator J1724$-$1443 was first imaged in \textsc{Difmap} \citep{Shepherd1994}, using hybrid mapping involving iterations of \textsc{clean} deconvolution and self-calibration. The antenna-based gain corrections obtained in \textsc{Difmap} were applied to the visibility data in AIPS. Then global fringe-fitting was repeated, now also considering the calibrator image, to account for the phases and delays introduced by the calibrator source structure. The AIPS task \textsc{calib} on J1724$-$1443 was also run with a solution interval of 2~min to further correct the amplitudes. Finally, the calibrated data file for Swift~J1727.8$-$1613 was produced by averaging the visibility data over all available spectral channels within the respective IFs, for subsequent analysis. 

\section{Results} 
\label{sec:resul}

\begin{figure}
    \centering
    \includegraphics[width=0.95\linewidth]{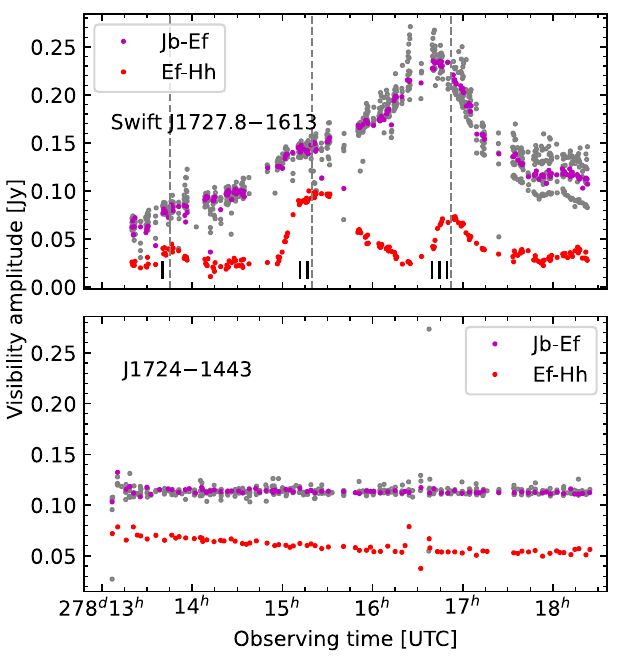}
    \caption{Visibility amplitude versus observing time. The three vertical dashed lines in the upper panel indicate the approximate peak positions of the visibility amplitudes of Swift~J1727.8$-$1613 on the Hh baselines (red curve). For the convenience of display, the plot is made with the baselines to the most sensitive Ef antenna, and only one IF (IF 5 at 4.927~GHz) is shown, with the data points further averaged over every 15 visibilities. The amplitudes measured on the Jb--Ef baseline are highlighted with magenta, data points on baselines between Ef and other European antennas are shown in gray. For comparison, the lower panel shows the nearly constant visibility amplitudes measured for the compact phase-reference calibrator quasar J1724$-$1443.}
    \label{fig:2}
\end{figure}

As shown in the upper panel of Fig.~\ref{fig:2}, prominent flux-density variation of the target source is evident within the 6-h observing interval with the EVN. Correlated flux densities (i.e., visibility amplitudes) measured on the baselines between the European antennas, and the baselines from Europe to Hh (South Africa) exhibit completely different behaviors. The amplitudes on the intra-European baselines overlap, except during the last hour, when the amplitudes on different baselines follow clearly different tracks. This implies that the source was initially unresolved at this scale, then became gradually resolved after the amplitude peak. 

The amplitudes on the Hh baselines are lower than those on the intra-European baselines throughout the entire observation period (Fig.~\ref{fig:2}). Notably, three peaks are visible (the first one can be tentatively identified). These facts indicate that the source had an extended emission feature and underwent significant structural changes during the observations, which were resolvable on the intercontinental baselines to Hh. 

For comparison, the amplitudes of the phase-reference quasar J1724$-$1443 on different baselines remain quite stable during the observing time (see the lower panel of Fig.~\ref{fig:2}). It confirms that the amplitude variation observed in the target source is not an instrumental effect but real. Note that the T6 antenna in China only observed a few scans at the beginning of the experiment, before Ef joined the array. T6 data are thus not included in Fig.~\ref{fig:2}.

\begin{figure*}
    \centering
    \includegraphics[width=1.0\linewidth]{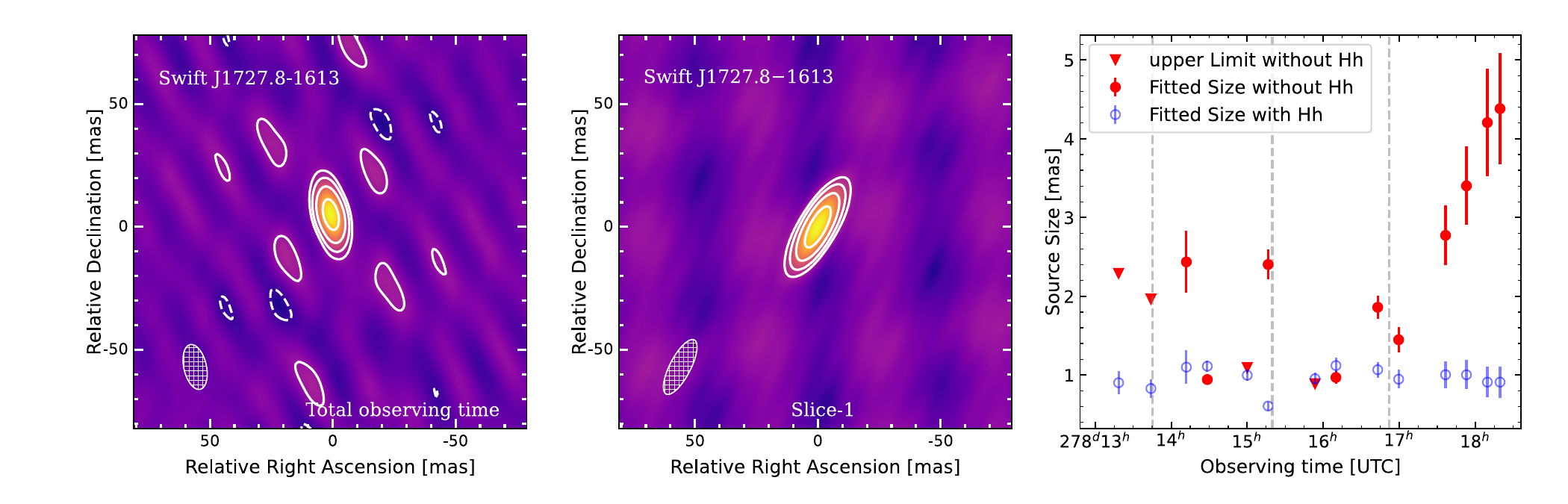}
    \caption{The \textsc{modelfit} images of Swift~J1727.8$-$1613 and the fitted angular size versus observing time. The two images were made using intra-European baselines only. The first contours are at $\pm5\sigma$ (left panel) and $\pm3\sigma$ (middle panel) image rms noise levels, respectively, and the positive contours increase by a factor of 2. The peak brightness and image rms noise level values are $138.8$ and $2.6$~mJy\,beam$^{-1}$ for the left panel, and $65.4$ and $1.8$~mJy\,beam$^{-1}$ for the middle panel, respectively. In the right panel, the red symbols represent the fitted sizes measured using only the European baselines, and the blue open circles are the ones measured when including Hh. The three vertical dashed lines are the same as in Fig.~\ref{fig:2}.}
    \label{fig:3}
\end{figure*}

Considering the inferred complex structure of the source and the absence of intermediate-length baselines in the array (see Fig.~\ref{fig:1}), we excluded Hh and T6, and only retained the European antennas for imaging. First, the entire phase-referenced dataset was fitted with one single circular Gaussian brightness distribution model component using the \textsc{Difmap} \textsc{modelfit} program. 
To improve the dynamic range, self-calibration (first phase-only, then amplitude and phase) was performed. As shown in the left panel of Fig.~\ref{fig:3}, no emission structure is obviously detected apart from a single component. Additionally, the source does not deviate much from the image phase center, indicating that the a priori coordinates for the antenna pointing were fairly accurate. There are some artifacts above $5\sigma$ image noise level around the source, which should be due to the flux density variation during the observing time rather than real features.

As the source was reasonably bright, we divided the data into 14 slices, each consisting of 2 or 3 consecutive scans ($10-20$~min) so as to investigate the possible structural changes \citep[cf.][]{Egron2017}. In each individual slice, a single circular Gaussian model component was fitted to the visibility data (this is justified by the fact that the source was unresolved initially and only tentatively resolved at the end of the EVN experiment, as seen by the European baselines). This time, only phase self-calibration was performed. The image made from the first slice is shown in the middle panel of Fig.~\ref{fig:3} as an example. The source does appear to expand after the visibility amplitude peak (see the right panel in Fig.~\ref{fig:3}). Intriguingly, the fitted size of the Gaussian model component seems to oscillate between 1 and 3~mas (see the discussion on this point in Section~\ref{subsec:simul}).

We also included Hh and repeated the model fitting. The fitted size appears stable (see right panel, Fig.~\ref{fig:3}). Notably, near the time of the second amplitude peak detected by the Hh baselines, the size drops. This can be attributed to the highest amplitudes observed by the Hh baselines at this time. All the fits appear satisfactory; however, due to the absence of intermediate baselines, these fits cannot represent the real structure for this complex source. Nevertheless, they do indicate the minimum extension of the source ($\sim 1$~mas), which is in the east--west direction \citep{Wood2024}. Note that this could only be considered as an indicative value due to the non-ideal $(u, v)$ coverage as shown in Fig.~\ref{fig:1}.

\begin{figure}
    \centering
    \includegraphics[width=0.9\linewidth]{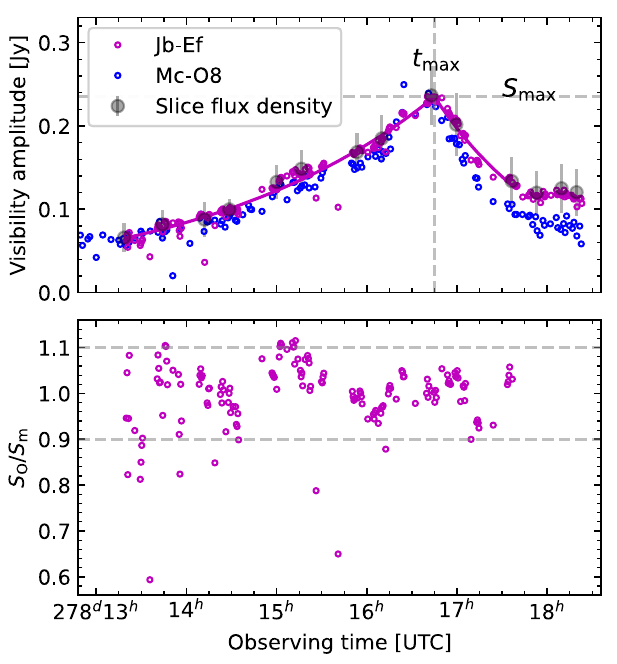}
    \caption{The rise and fall timescales of the visibility amplitude. In the upper panel, the open circles are visibility amplitudes from IF 5, with the data further averaged over every 15 and 30 visibilities for the Jb--Ef (magenta) and Mc--O8 (blue) baselines, respectively, to improve the signal-to-noise ratio. The magenta curve indicates the broken exponential function fitted to the Jb--Ef visibility amplitudes. The source flux densities of the 14 slices, obtained by the \textsc{modelfit} program in \textsc{Difmap} (only European baselines), are shown with gray filled circles. The lower panel displays the ratio of the observed visibility amplitudes ($S_{\rm O}$) to those calculated by the exponential model ($S_{\rm m}$).}
    \label{fig:4}
\end{figure}

The measured flux densities in the 14 slices (Fig.~\ref{fig:4}, European baselines only) align well with the visibility amplitude evolution of the short Jb--Ef baseline, confirming the reliability of the \textsc{modelfit} results. The source parameters are presented in Tables~\ref{tab:1} and \ref{tab:2} in the Appendix.

\section{Discussion} 
\label{sec:disc}

\subsection{Jet Kinematics and Energetics}
\label{subsec:JK}

The variability feature observed on the Hh baselines (see Fig.~\ref{fig:2}) suggests that the source has structure in the north--south direction, since this is the orientation of the baselines from the European antennas to Hh (Fig.~\ref{fig:1}). It is consistent with previous VLBA/LBA observations during the hard intermediate state (HIS), which revealed a prominent jet extending to the south \citep{Wood2024}. That the amplitudes on Hh baselines are rising and declining alternately can be naturally explained with a plasma blob ejected from the radio core, which dominates the major flare and moves relativistically to the south. In this case, the angular distance $\Theta$ of the blob positions corresponding to two adjacent peaks ($\sim 1.5$~h apart, see the upper panel of Fig.~\ref{fig:2}) can be estimated with $\lambda/B$
(here, $\Theta$ is in radians, $\lambda$ is the observing wavelength in meters, and $B$ is the projected baseline length in meters). Therefore, the proper motion of the blob on the sky would be $\sim30$~mas\,d$^{-1}$ (for this estimation, the average projected baseline length $B = 110\,{\rm M}\lambda$ is taken; see Fig.~\ref{fig:1}). Taking into account the source distance \citep{Mata2025}, this would correspond to an apparent speed of $v_\mathrm{app} \sim 0.6\,c$ (where $c$ denotes the speed of light).

The X-ray polarization measurements suggest that the jet viewing angle is {\bf $\theta_{\rm V}\sim(30-50)^\circ$ \citep{Svoboda2024}}. This, together with the apparent speed ($0.6\,c$), would yield an intrinsic speed of the ejected blob of $v\sim(0.5-0.6)\,c$. The corresponding Doppler factor would be $\delta\sim1.3-1.6$.

Considering the large viewing angle, another possibility cannot be ruled out: the behavior of the visibility amplitude variation seen on the Hh baselines may be caused by two visible ejected blobs, moving symmetrically and in opposite directions away from the core. For this case, the apparent speed of $\sim 0.6\,c$ should be seen as the apparent separation speed of the two blobs. Thus, the estimated blob speed and Doppler factor would be $v\sim(0.4-0.5)\,c$ and $\delta\sim1.2-1.5$, respectively, both of which are slightly lower than those in the one-blob scenario.

In both scenarios, a larger viewing angle corresponds to a smaller jet speed and a smaller Doppler factor. The proper motion of $\sim 1.25$~mas\,h$^{-1}$ measured by our EVN experiment is broadly consistent with those measured by \citet{Wood2024,Wood2025}, i.e., $\sim(0.6-3)$~mas\,h$^{-1}$, although the source was in different spectral states. Using the proper motion of the fastest-moving component of $3.18\pm0.10$~mas\,h$^{-1}$ and the source distance of $3.4\pm0.3$\,kpc, \citet{Wood2025} constrained the source viewing angle to $\theta_{\rm V} < 67^{\circ}$, which agrees with the X-ray polarization measurements \citep{Svoboda2024}.

As shown by the last four red data points in the right panel of Fig.~\ref{fig:3}, where a single Gaussian model component was used to fit the entire source, the source exhibited linear expansion with an apparent angular expansion speed of $\sim57 \pm 23$~mas\,d$^{-1}$. This value is somewhat higher, while it can be consistent with the previously measured proper motion of $\sim 30$~mas\,d$^{-1}$, using the three amplitude peaks seen on Hh baselines, when considering the measurement uncertainties. If the jet proper motion remained constant throughout the EVN observation, the minor flare observed during the final hour of the experiment could be associated with a new smaller ejection event from the radio core. However, if the jet did undergo acceleration. The minor flare could instead be attributed to an internal shock, produced by the newly ejected blob catching up and colliding with a previously ejected one.

%This is apparently faster, implying the jet likely underwent acceleration. This could happen if, for instance, the minor flare observed in the last one hour of the experiment was caused by an inner shock, produced by the newly ejected blob catching up and colliding with the previous ejected blob. For the one-blob scenario, the jet speed would be $v\sim 0.7\,c$. If, however, we assume that the source expanded symmetrically, corresponding to the two visible blobs scenario, the jet speed {\bf would be $v\sim(0.6-0.7)\,c$}. It is also possible that the apparent expansion angular speed ($\sim50$~mas\,d$^{-1}$) is overestimated. In this case, the minor flare could be associated with a new minor ejection event from the radio core.

As shown in Fig.~\ref{fig:4}, the visibility amplitude evolves quite smoothly during the EVN experiment, except during the last hour, when the source may undergo a new minor flare. The smooth characteristic implies that the Doppler factor does not change significantly (otherwise, we would expect large, unsmoothed amplitude variation, which is not observed). We fitted the amplitude evolution versus time on the Jb--Ef baseline with a broken exponential function to derive the rise and fall timescales of the major flare \citep{Valtaoja1999}:
\begin{equation}
S_{\rm corr} = 
    \begin{cases}
        S_{\rm max} \times \exp\left(\frac{t-t_{\rm max}}{\tau_{\rm rise}}\right), & t < t_{\rm max} \\
        S_{\rm max} \times \exp\left(\frac{t_{\rm max}-t}{\tau_{\rm fall}}\right), & t > t_{\rm max}
    \end{cases}
    \label{eq:1}
\end{equation}
Here, $S_{\rm corr}$ is the correlated flux density (i.e., visibility amplitude). The amplitude peak $S_{\rm max}$ (235~mJy) and the corresponding time $t_{\rm max}$ (16:45) are visually identified from the upper panel of Fig.~\ref{fig:4}.
The fall timescale ($\tau_{\rm fall} = 0.059 \pm 0.002$~d) is shorter than the rise timescale 
($\tau_{\rm rise} = 0.112 \pm 0.001$~d). This is different from what is often observed in the flares of active galactic nuclei, where the fall timescale is usually longer \citep{Valtaoja1999}. This difference is partly due to the source becoming resolved after the amplitude starts to decrease. As shown in Fig.~\ref{fig:4}, the visibility amplitudes on the baseline Mc--O8 (the longest European baseline) are lower than those on the short Jb--Ef baseline after the amplitude peak, indicating that the source is indeed resolved.

Measuring the volume of an ejected blob is challenging due to its rapidly evolving structure. \citet{Fender2019} parameterized the total energy of a blob at the peak of its light curve in terms of its effective expansion speed, defined as $\beta_{\rm e} = v_{\rm e}/c$. At the point of minimum energy, the magnetic field energy $E_{\rm B}$ and the particle energy $E_{\rm p}$ satisfy $E_{\rm B} = (6/11)\,E_{\rm p}$, which differs slightly from the equipartition condition: $E_{\rm B} = (3/4)\,E_{\rm p}$ \citep{Longair2011}.

Assuming that the approaching blob dominated the major flare observed by the EVN, its flux-density evolution can be approximately described by Eq.~\ref{eq:1}. \citet{Fender2019} provided a method to estimate the blob parameters using a single-frequency observation. For relativistic cases, Doppler correction is required. With the estimated Doppler factor of $\delta \sim 1.2–1.6$, we obtained the blob brightness temperature of $T_{\rm in} \sim 5 \times 10^{10}$~K, effective expansion speed of $\beta_{\rm e} = 0.03–0.04$, minimum energy of $E_{\rm in} = (1.4–3.3) \times 10^{38}$~erg, and magnetic field strength of $B_{\rm in} = (0.4–0.5)$~G, in the rest frame of the approaching blob. The blob’s kinetic energy is given by $E_{\rm k} = (\Gamma - 1)\,E_{\rm in}$, which leads to a kinetic power of $P = E_{\rm k} / \tau_{\rm rise} \sim 10^{33}$\,erg\,s$^{-1}$ (here, the bulk Lorentz factor $\Gamma \sim 1.1-1.3$). These parameters are broadly consistent with those estimated for the four flare cases in \citet{Fender2019}. Notably, the brightness temperature $T_{\rm in}$ is in good agreement with the equipartition value of $5\times 10^{10}$~K suggested by \citet{Readhead1994}, while the blob’s expansion speed $\beta_{\rm e}$ is much less than 1. This suggests that treating the rise timescale $\tau_{\rm rise}$ as the light-crossing time (i.e., assuming the blob expansion speed equals $c$), will significantly overestimate the blob size. The soft X-ray photon flux ($2-4$~keV), measured simultaneously with the EVN observations, is approximately $15$\,ph\,cm$^{-2}$\,s$^{-1}$ (see Fig.~\ref{fig:7}). Assuming the radiation efficiency $\eta=0.1$, the estimated source accretion power would be $P_{\rm acc} = L_{\rm disk}/\eta=\dot{\rm M}\,c^2 \sim 10^{39}$~erg\,s$^{-1}$ ($L_{\rm disk}$ is the disk luminosity and $\dot{\rm M}$ the accretion rate), which is dramatically higher than the blob kinetic power. If the BH has a mass of 10~${\rm M}_\sun$, the $L_{\rm disk}$ will be below the Eddington luminosity, and thus the system will be in the sub-Eddington accretion regime.

%Note that the estimates of the jet parameters rely on the source distance. At the moment, the accurate parallax distance for the source is not available; thus, the distance-dependent jet parameters presented here are only indicative. Recently, a new distance measurement for Swift~J1727.8--1613 was reported, suggesting that the source is likely located farther away, at $5.5^{+1.4}_{-1.1}$~kpc \citep{Burridge2025}. At this distance, the estimated proper motion of $\sim 30$~mas\,${\rm d}^{-1}$ will correspond to an apparent speed of $v_{\rm app}\sim 1.0\,c$. The apparent expansion angular speed of $\sim50$~mas\,d$^{-1}$ will instead correspond to $\sim 1.6\,c$.

\subsection{Simulation}
\label{subsec:simul}

\begin{figure*}
    \centering
    \includegraphics[width=1.0\linewidth]{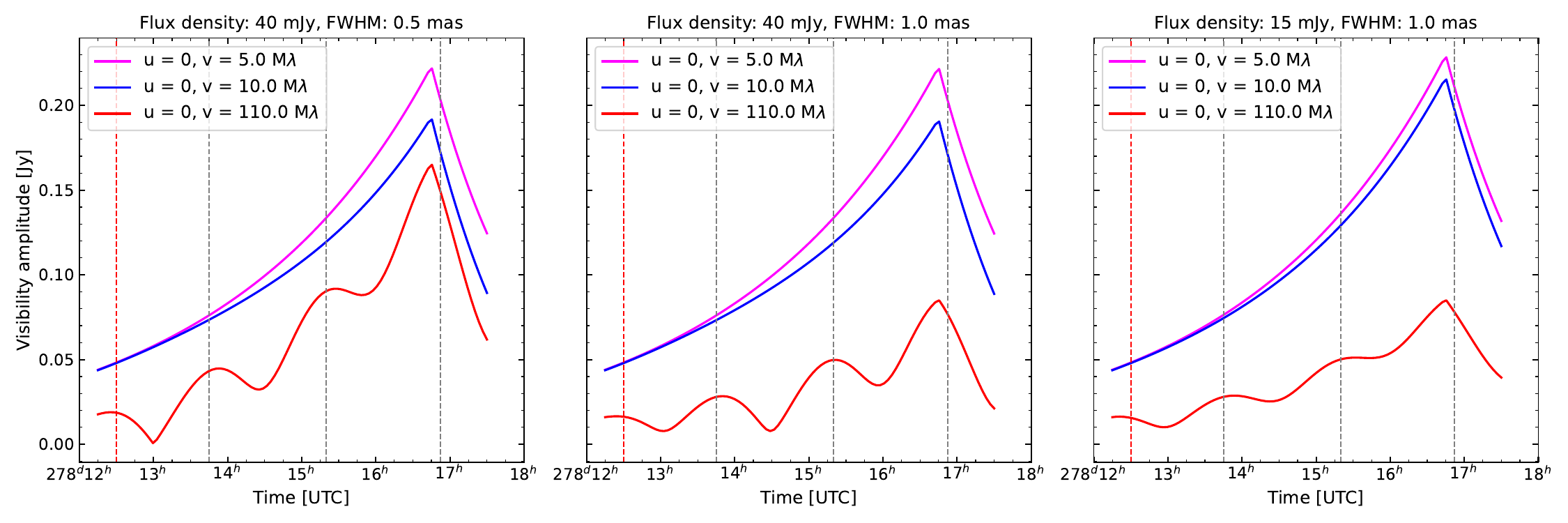}
    \caption{The simulated evolution of the visibility amplitude over time. The amplitudes at three points in the visibility plane, i.e., (0, 5)~M$\lambda$, (0, 10)~M$\lambda$, and (0, 110)~M$\lambda$, are selected to represent two short baselines and one long baseline. The model consists of two components. The flux density of the stationary component and the FWHM diameter of the moving component are 40~mJy and 0.5~mas for the left panel, 40~mJy and 1.0~mas for the middle panel, and 15~mJy and 1.0~mas for the right panel, respectively (see the text for further details). The start time of the EVN observation is indicated by the vertical red dashed line. The three vertical gray dashed lines are the same as those in Fig.~\ref{fig:2}.}
    \label{fig:5}
\end{figure*}

\begin{figure*}
    \centering
    \includegraphics[width=1.0\linewidth]{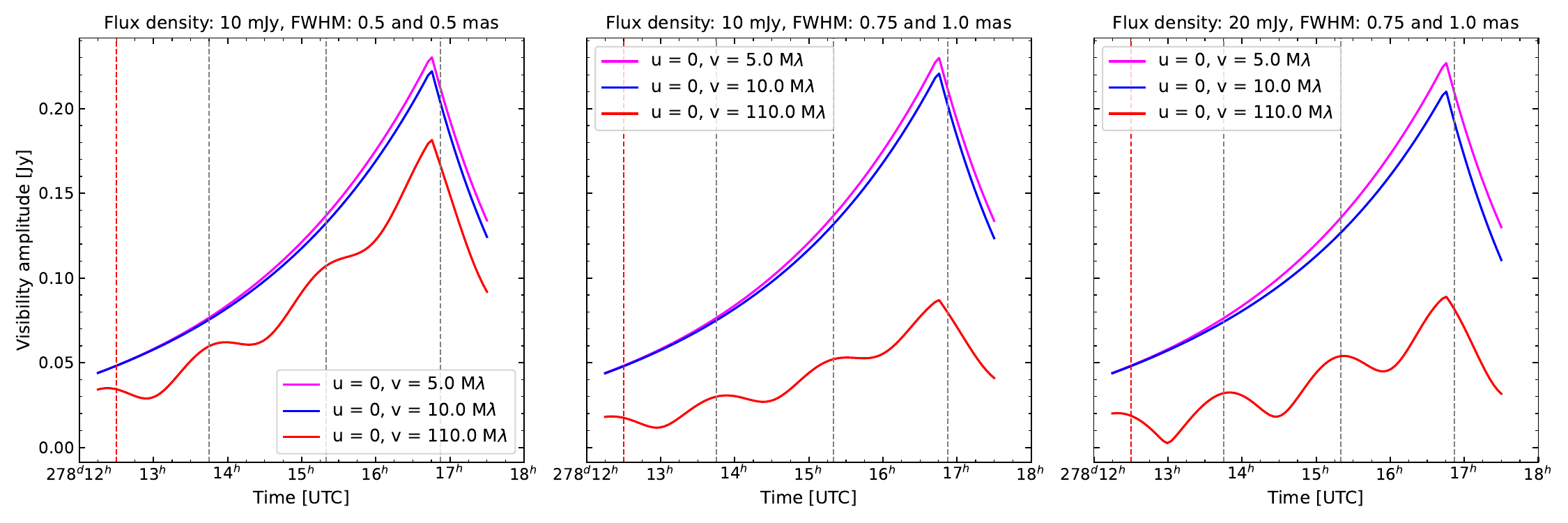}
    \caption{The simulated evolution of the visibility amplitude over time. The amplitudes at three points in the visibility plane, i.e., (0, 5)~M$\lambda$, (0, 10)~M$\lambda$, and (0, 110)~M$\lambda$, are selected to represent two short baselines and one long baseline. The model consists of two components. The flux density of the receding component and the FWHM diameters of the receding and approaching components are 10~mJy, 0.5~mas and 0.5~mas for the left panel, 10~mJy, 0.75~mas and 1.0~mas for the middle panel, and 20~mJy, 0.75~mas and 1.0~mas for the right panel, respectively (see the text for further details). The start time of the EVN observation is indicated by the vertical red dashed line. The three vertical gray dashed lines are the same as those in Fig.~\ref{fig:2}.}
    \label{fig:6}
\end{figure*}

We conducted a simulation to determine whether the two scenarios proposed in Section~\ref{subsec:JK} could indeed explain the observed phenomena shown in the upper panel of Fig.~\ref{fig:2}. We first investigated the one-blob scenario, in which two circular Gaussian brightness distribution model components were considered: one, with a full width at half maximum (FWHM) of 1~mas, remains stationary at the image phase center, while the other moves south at an apparent angular speed of 30~mas~d$^{-1}$. The flux density of the stationary component remains stable, and the sum of the flux densities of the two components is described by Eq.~\ref{eq:1}. The rise and fall timescales, as well as the amplitude peak value, are set to the fitted values (i.e., $\tau_{\rm rise} = 0.112$~days, $\tau_{\rm fall} = 0.059$~days, and $S_{\rm max} = 235$~mJy). The simulation begins at the inferred ejection time (12:15), when the two components are superposed, and ends at 17:30, when the minor flare (i.e., the small bump observed in the amplitude evolution over time at the end of the EVN observation; see the upper panel of Fig.~\ref{fig:2}) begins to appear.

The visibility amplitudes were calculated following the van Cittert--Zernike theorem. Initially, we set the flux density of the stationary component to 40~mJy and the FWHM of the moving component to 0.5~mas (only these two parameters are changeable for the one-blob case). The evolution of the visibility amplitude over time is shown in the left panel of Fig.~\ref{fig:5}. It is evident that the amplitudes on the short baselines first rise and then fall, while the amplitude on the long baseline exhibits oscillation. This behavior is roughly consistent with the measurements. However, on the long baseline, the third peak appears to be the highest, and on the short baselines, the source appears to be resolved (well) before the amplitude peak, which is at odds with what was observed.

To explore this further, we increased the FWHM of the moving component to 1.0~mas and repeated the calculation. As shown in the middle panel of Fig.~\ref{fig:5}, the visibility amplitude on the long baseline is suppressed as expected. However, the amplitude evolution of the short baselines remains unchanged. We then reduced the flux density of the stationary component to 15~mJy. This time, as shown in the right panel of Fig.~\ref{fig:5}, while the amplitude oscillation on the long baseline becomes weaker, the amplitudes on the short baselines shift towards overlapping.

From the simulation, we can conjecture that the stationary component (i.e., the radio core) could initially be brighter than the moving component, as the latter is expected to be optically thick at the beginning. As the flux density of the moving component and the angular distance between the two components increase, strong oscillation appears on the long baseline. The stationary component gradually weakens, causing the amplitudes on the short baselines to remain in an overlapping state. During the same time, the size of the moving component continuously expands, resulting in a relatively lower amplitude of the third peak on the long baseline. Therefore, the simulation indicates that the one-blob scenario is viable and provides additional insights into the evolution of the source structure and flux density. 

Since interference depends on the angular separation of the two components rather than their absolute positions, this simulation is also applied to the two-blob scenario. In this case, the counter-blob (i.e., the receding component in the north) can be treated as the stationary component. What is different here is that the flux densities and FWHMs of the two components are expected to be closely related if the bipolar jet is symmetric.

The flux-density ratio of the approaching component ($S_{\rm a}$) to the receding component ($S_{\rm r}$) can be calculated as \citep{Miller-Jones2004}:
\begin{equation}
    \frac{S_{\rm a}}{S_{\rm r}} = \left(\frac{1+\beta \cos \theta_{\rm V}}{1-\beta \cos \theta_{\rm V}}\right)^{3-\alpha}
    \label{eq:2}
\end{equation}
Here, $\beta = v/c$. Taking into account the jet parameter estimates in Section~\ref{subsec:JK} and assuming a spectral index of $\alpha = 0.5$ (here, the flux density $S_\nu \propto \nu^\alpha$), the flux-density ratio of the receding component to the total flux density of the source would be $\sim 0.1-0.2$. A higher ratio corresponds to a larger viewing angle. Since there is a delay between the observed flux densities of the two components, this flux-density ratio is only expected to be true at the ejection time. The delay can be calculated as \citep{Miller-Jones2004}:
\begin{equation}
    t_{\rm r} = \left(\frac{1-\beta \cos \theta_{\rm V}}{1+\beta \cos \theta_{\rm V}}\right) t_{\rm a}
    \label{eq:3}
\end{equation}
Here, both $t_{\rm r}$ and $t_{\rm a}$ are counting from the ejection time. The radiation from the approaching component, emitted at $t_{\rm a}$, is in fact observed simultaneously with that from the receding component, emitted at $t_{\rm r}$. Using the jet parameters estimated in Section~\ref{subsec:JK}, we can obtain $t_{\rm r} = (0.4-0.6)\,t_{\rm a}$. This delay
causes the observed flux density and angular size (i.e., FWHM) of the receding component to evolve very slowly. Considering these factors, we started the simulation with a low flux density of 10~mJy for the receding component and allowed the FWHMs of both components to be changeable, when calculating the visibility amplitude evolution over time (see Fig.~\ref{fig:6}). The similarity of the simulated results shown in Fig.~\ref{fig:5} and \ref{fig:6} suggests that the two-blob scenario is also a viable explanation for the data.

At first glance, the apparent oscillation of the fitted size (right panel in Fig.~\ref{fig:3}) may suggest that large changes occurred in the jet direction. To make a rough estimation, the ratio of the observed flux density $S_{\rm O}$ to the flux density calculated by the exponential model $S_{\rm m}$ is:
\begin{equation}
    \frac{S_{\rm O}}{S_{\rm m}} = \left(1+\frac{\Delta\delta}{\delta}\right)^{3-\alpha} 
    \label{eq:4}
\end{equation}
From the lower panel of Fig.~\ref{fig:4}, we can see $0.9<S_{\rm O}/S_{\rm m}<1.1$. Assuming the spectral index $\alpha = 0.5$, the variation range of the Doppler factor $|\Delta\delta|<\delta\times5\%$. The apparent size oscillation could be partly due to the wobbling of the jet. However, the small variation range of the Doppler factor argues against jet wobbling as the main reason. In fact, considering the $\sim10$\% systematic error in the visibility amplitude measurements, the ``pattern'' seen in the lower panel of Fig.~\ref{fig:4} does not support that changes in the Doppler factor occurred. From Fig.~\ref{fig:5} and \ref{fig:6}, we can see the amplitude evolution over time on short baselines is not sensitive to changes in the component sizes but rather to changes in the flux densities of the two components. Due to interference, when the flux densities of the north and/or south components change (e.g., resulting from opacity variations), the amplitudes on different short baselines will either shift towards overlapping or diverge, leading to a smaller or larger fitted size. This suggests that the apparent oscillation of the fitted size could be attributed to interference of the radiation from the two components. The linear increase of the measured size, seen in the last hour (right panel of Fig.~\ref{fig:3}), indicates that the size measurements are dominated by the separation of the two components. Therefore, in this case, the measured sizes can be used to assess the change of the source size. 

For comparison, during the 2015 outburst of V404~Cygni, \citet{Miller-Jones2019} observed significant changes in the ejection directions of different blobs, which, however, were not seen in Swift~J1727.8--1613 \citep{Wood2025}. A key difference between the two sources is their accretion states. V404~Cygni was in super-Eddington accretion regime, and the changes in jet direction were explained as being driven by the precessing, misaligned, and puffed-up slim disk \citep{Miller-Jones2019}.

\subsection{Disk--Jet Correlation}
\label{subsec:DJC}

\begin{figure}
    \centering
    \includegraphics[width=1.0\linewidth]{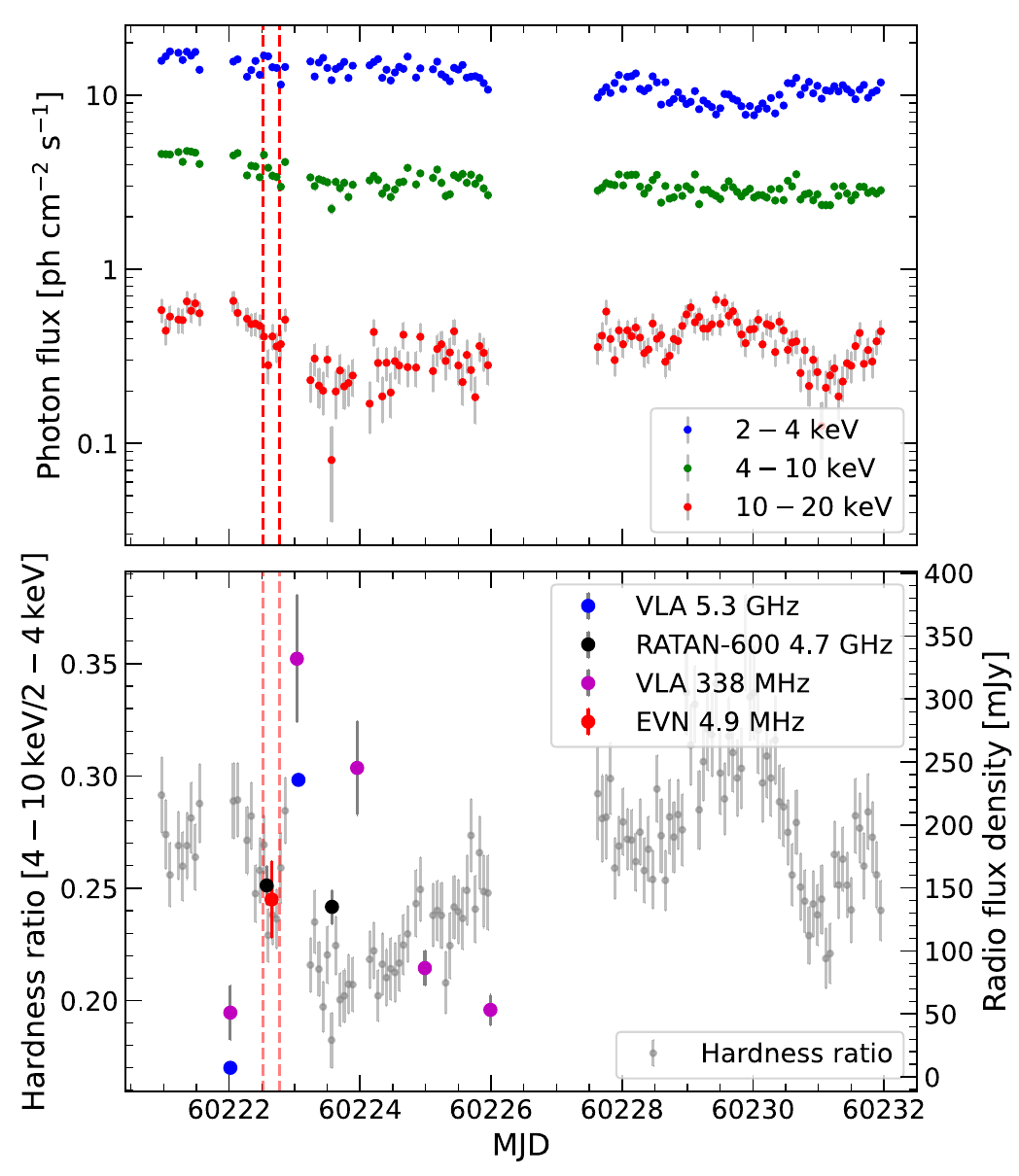}
    \caption{The X-ray photon flux, radio flux density, and X-ray hardness variations over time. The beginning and the end of the EVN observing period are shown by the two vertical red dashed lines. The X-ray data are taken from the MAXI archive \citep{Matsuoka2009}. The radio flux densities at $338$~MHz, $4.7$~GHz, and $5.3$~GHz are taken from \citet{Peters2023}, \citet{Trushkin2023}, and \citet{Miller-Jones2023a}, respectively.}
    \label{fig:7}
\end{figure}

The X-ray and radio monitoring of this source allows us to explore the disk--jet correlation \citep{Matsuoka2009, Peters2023}. Fig.~\ref{fig:7} shows that the EVN experiment was conducted right during the rising phase of a large radio flare lasting for 4~days \citep{Peters2023}. The flux density $141\pm31$~mJy, measured with the EVN (only European baselines and the entire observing time) agrees well with that measured by the RATAN-600 radio telescope \citep[][lower panel in Fig.~\ref{fig:7}]{Trushkin2023}, confirming the accuracy of the EVN amplitude calibration. During the 6-h EVN observation, a radio flare accompanied by ejected plasma blob(s) was observed, which suggests that such flare/ejection events likely occurred quite frequently throughout the entire 4-day-long large radio flare. This further implies that, if the flare timescale is measured using less frequent radio monitoring, the jet power \citep{Fender2006,Carotenuto2021} may be (severely) underestimated. This may partly explain the inconsistency in the estimates of the jet power of MAXI~J1820+070 \citep{Bright2020}.

It can be seen that the radio flare illustrated in Fig.~\ref{fig:7} is accompanied by a softening of the X-ray spectrum, primarily caused by the weakening of the hard X-ray emission. This indicates that some changes, such as disturbances of the magnetic field in the inner hot disk, triggered the hard X-ray flux variation and the radio flare. However, the hot disk did not collapse completely, suggesting that the radio core remained active throughout. An interesting phenomenon is that the radio quenching, detected by the VLA \citep{Miller-Jones2023a}, occurred about half a day before the hard X-ray flux began to decline, and about $1.5$~days before the hard X-ray flux reached its lowest level. This suggests that the disturbances likely originated at the inner edge of the hot accretion disk (close to the BH). The X-ray polarization observations can constrain the geometry of the corona, which, in hard state, is the hot accretion disk in the truncated disk model \citep{Eardley1975,Ichimaru1977}, and thus help to test/refine the model. However, no IXPE observations were available during the period of hard X-ray flux decline in which our EVN experiment took place \citep[][]{Ingram2024}.

The type-C QPOs were always observed during the monitoring campaign carried out by the Hard X-ray Modulation Telescope (HXMT), which ceased on October 4, 2023, due to solar obscuration \citep{Yu2024}. At the end of this monitoring, the QPO frequency reached $\sim 7$~Hz. During 03:00--11:00 UTC on October 5, 2023, just prior to our EVN observation, the dynamic power spectrum observed by NICER showed that the 0.1--10~Hz fractional rms at the 4--10~keV band dropped from 10\% to 4\%, and the 7-Hz type-C QPO became undetectable \citep{Bollemeijer2023}. These timing characteristics and the subsequent occurrence of the ejection events are consistent with the transient jet ejection picture established for BH LMXBs \citep{Miller-Jones2012}. The presence of high-level hard X-ray flux ($>0.5$~Crab at the 10--20~keV band) suggests that the source should remain in the HIS. Unlike, for instance, MAXI~J1820+070 \citep{You2023}, the source did not transition into the soft state abruptly. Before finally evolving into the soft state, the source underwent a sequence of X-ray/radio flares starting from September 11, 2023 \citep{Yu2024,Podgorny2024}, presumably due to a gradual release of the accumulated magnetic field energy in the central engine.

\section{Summary} 
\label{sec:sum}

A large-amplitude flare was observed in the BH LMBXB Swift~J1727.8$-$1613 during our 5-GHz VLBI observations on October 5, 2023. It is inferred to be caused by an ejection event. Our findings are summarized as follows:
\begin{itemize}
    \item From the evolution of visibility amplitude over time, it can be seen that the source was initially unresolved and only tentatively resolved at the end of the EVN observation on the European baselines, while the visibility amplitudes on the Hh baselines exhibited oscillation, indicating the source was resolvable on the longest baselines. These behaviors can be interpreted as resulting from either a single relativistic plasma blob ejected from the stationary radio core (i.e., the one-blob scenario), or two blobs moving symmetrically away from the core in opposite directions (i.e., the two-blob scenario).
    \item Using the oscillation characteristic of the visibility amplitudes on Hh baselines, we estimated the angular separation speed of the two components to be 30~mas\,d$^{-1}$. Taking the distance measurement of $\sim3.4$~kpc \citep{Mata2025} and the view angle estimate of $(30-50)^\circ$ \citep{Svoboda2024}, for the one-blob scenario into account, we estimated the blob speed and the Doppler factor as $v\sim(0.5-0.6)\,c$ and $\delta \sim 1.3-1.6$, respectively. For the two-blob case, the obtained values are slightly lower: $v\sim(0.4-0.5)\,c$ and $\delta \sim 1.2-1.5$. Both scenarios can be supported by the simulation. 
    \item We divided the data into 14 slices and fitted one single Gaussian model component to each slice. For the case of including only the European baselines, initially, the fitted size exhibited oscillation, which can be attributed to interference between the radiation from two components. At the end of the EVN observation (i.e., the last hour), the fitted size showed linear expansion, indicating that the measured size was dominated by the angular separation of the two components.
    \item Assuming the approaching blob dominated the major flare seen by the EVN, we estimated the jet kinetic power of $P \sim 10^{33}$~erg\,s$^{-1}$, which is much lower than the accretion power of $P_{\rm acc} = \dot{\rm M}\,c^2 \sim 10^{39}$~erg\,s$^{-1}$. Changes in the hard X-ray flux suggest that the ejection event was related to the (magnetic) activities in the hot disk. 
\end{itemize}

%By analyzing the visibility amplitude evolution over time, we estimated the proper motion of the blob(s). Taking the distance measurement of $\sim3.4$~kpc, for the one-blob scenario, we estimated blob speed to be $v\sim(0.5-0.6)\,c$. For the two-blob case, {\bf the obtained speed is slightly lower, $v\sim(0.4-0.5)\,c$}. Both scenarios can be supported by the simulation. Initially, the jet exhibited apparent size oscillation, which could be attributed to the interference effect. The flare was inferred to be dominated by the approaching blob, and when it became transparent, the source appeared to expand linearly in size, {\bf with a speed of $v\sim(0.6-0.7)\,c$}. % (assuming that the jet expansions on either side of the core were symmetric). 
%{\bf Assuming that the blob expands at the speed of light will significantly overestimate its size, and thus lead to an underestimation of the variability brightness temperature, and consequently, the variability Doppler factor.} Changes in the hard X-ray flux suggest that the ejection event was related to the (magnetic) activities in the hot disk. 

Bright Galactic X-ray binaries like Swift~J1727.8$-$1613 offer valuable opportunities for the short-timescale exploration of the jet activities (flux density and structural changes) with VLBI. Combined with multi-wavelength monitoring, especially in the X-ray band, the VLBI observations can provide critical insights into the disk--jet coupling and help draw a more complete picture of accretion and ejection process.

%% IMPORTANT! The old "\acknowledgment" command has be depreciated. It was
%% not robust enough to handle our new dual anonymous review requirements and
%% thus been replaced with the acknowledgment environment. If you try to 
%% compile with \acknowledgment you will get an error print to the screen
%% and in the compiled pdf.
%% 
%% Also note that the akcnowlodgment environment does not support long amounts of text. If you have a lot of people and institutions to acknowledge, do not use this command. Instead, create a new \section{Acknowledgments}.
\begin{acknowledgments}
The EVN is a joint facility of independent European, African, Asian, and North American radio astronomy institutes. Scientific results from data presented in this publication are derived from the following EVN project code: RM019. This research has made use of MAXI data provided by RIKEN, JAXA and the MAXI team. H.M.C. acknowledges support from the Hebei Natural Science Foundation of China (grant no. A2022408002), and the Basic Research Plan for Key Research Projects of Higher Education Institutions in Henan Province (Project no. 25A160001). S.F. and K.\'E.G. were supported by the Hungarian National Research, Development and Innovation Office (NKFIH OTKA grant K134213 and excellence grant TKP2021-NKTA-64).
K.\'E.G. was also supported by the HUN-REN.
\end{acknowledgments}

%% To help institutions obtain information on the effectiveness of their 
%% telescopes the AAS Journals has created a group of keywords for telescope 
%% facilities.
%
%% Following the acknowledgments section, use the following syntax and the
%% \facility{} or \facilities{} macros to list the keywords of facilities used 
%% in the research for the paper.  Each keyword is check against the master 
%% list during copy editing.  Individual instruments can be provided in 
%% parentheses, after the keyword, but they are not verified.

%\vspace{5mm}
%\facilities{HST(STIS), Swift(XRT and UVOT), AAVSO, CTIO:1.3m,
%CTIO:1.5m,CXO}

%% Similar to \facility{}, there is the optional \software command to allow 
%% authors a place to specify which programs were used during the creation of 
%% the manuscript. Authors should list each code and include either a
%% citation or url to the code inside ()s when available.

%\software{astropy \citep{2013A&A...558A..33A,2018AJ....156..123A},  
%          Cloudy \citep{2013RMxAA..49..137F}, 
%          Source Extractor \citep{1996A&AS..117..393B}
%          }

%% Appendix material should be preceded with a single \appendix command.
%% There should be a \section command for each appendix. Mark appendix
%% subsections with the same markup you use in the main body of the paper.

%% Each Appendix (indicated with \section) will be lettered A, B, C, etc.
%% The equation counter will reset when it encounters the \appendix
%% command and will number appendix equations (A1), (A2), etc. The
%% Figure and Table counter will not reset.

\appendix

\section{The source parameters of Swift J1727.8$-$1613}

The source flux densities and angular sizes determined for the 14 slices using the \textsc{modelfit} program in \textsc{Difmap} with only the intra-European baselines are presented in Table~\ref{tab:1}. Table~\ref{tab:2} presents the same measurements but including the Hh antenna as well. The uncertainties were estimated following \citet{Lee2008}, where $10\%$ of the measured flux density was added in quadrature to the thermal noise, to account for systematic amplitude calibration errors. When the fitted angular size is smaller than the resolution limit of the VLBI array, this limit is used as the upper limit for the angular size \citep{Kovalev2005}. 

\begin{table*}[h]
      \caption{The source parameters of Swift J1727.8$-$1613 with European-only baselines}
         \label{tab:1}
%         \begin{threeparttable}
         \centering
         \begin{tabular}{c c c c | c c c c}
           \hline
           \noalign{\smallskip}
           Slice & Time [d]  &$S_{\nu}$ [mJy]  & $\theta$ [mas] & Slice & Time [d] &  $S_{\nu}$ [mJy] & $\theta$ [mas] \\
%                 & &$b_{\rm max} \times b_{\rm min}$& & & & \\
           \noalign{\smallskip}
           \hline
           \noalign{\smallskip}
           1 & 0.55451 & 66 $\pm$17 & $<$2.29 & 8  & 0.67361  & 185$\pm$28 & 0.97$\pm$0.08 \\
           \noalign{\smallskip}
           2 & 0.57222 & 82 $\pm$18 & $<$1.97 & 9  & 0.69652  & 237$\pm$36 & 1.86$\pm$0.15 \\
           \noalign{\smallskip}
           3 & 0.59166 & 88 $\pm$22 & 2.44$\pm$0.39 & 10 & 0.70798  & 202$\pm$38 & 1.45$\pm$0.17 \\
           \noalign{\smallskip}
           4 & 0.60312 & 100$\pm$13 & 0.94$\pm$0.05 & 11 & 0.73368  & 134$\pm$29 & 2.77$\pm$0.38 \\
           \noalign{\smallskip}
           5 & 0.62500 & 133$\pm$21 & $<$1.10 & 12 &  0.74514 & 120$\pm$27 & 3.40$\pm$0.50 \\
           \noalign{\smallskip}
           6 & 0.63645 & 149$\pm$22 & 2.40$\pm$0.19 & 13 & 0.75660 & 125$\pm$30 & 4.21$\pm$0.68 \\
           \noalign{\smallskip}
           7 & 0.66215 & 169$\pm$23 & $<$0.89 &  14  & 0.76354  & 120$\pm$28 & 4.38$\pm$0.70 \\
           \hline
         \end{tabular}
         \begin{tablenotes}
         \item \emph{Notes}. The angular sizes and flux densities are shown in Fig.~\ref{fig:3} (right panel) and Fig.~\ref{fig:4}, respectively. Cols.~2 and 6 -- time on October 5, 2023 in fractions of a day (UTC); Cols. 3 and  7 –- modeled flux density; Cols. 4 and 8 –- fitted angular size (FWHM of the circular Gaussian brightness distribution model component). 
         \end{tablenotes}
 %        \end{threeparttable}
   \end{table*} 

\begin{table*}[]
      \caption{The source parameters of Swift J1727.8$-$1613 with European and Hh antennas}
         \label{tab:2}
%         \begin{threeparttable}
         \centering
         \begin{tabular}{c c c c | c c c c}
           \hline
           \noalign{\smallskip}
           Slice & Time [d]  &$S_{\nu}$ [mJy]  & $\theta$ [mas] & Slice & Time [d] &  $S_{\nu}$ [mJy] & $\theta$ [mas] \\
%                 & &$b_{\rm max} \times b_{\rm min}$& & & & \\
           \noalign{\smallskip}
           \hline
           \noalign{\smallskip}
           1 & 0.55451 & 66 $\pm$16 & 0.90$\pm$0.15 & 8  & 0.67361  & 186 $\pm$28 & 1.12$\pm$0.10 \\
           \noalign{\smallskip}
           2 & 0.57222 & 81 $\pm$18 & 0.83$\pm$0.12 & 9  & 0.69652  & 233$\pm$37 & 1.07$\pm$0.10 \\
           \noalign{\smallskip}
           3 & 0.59166 & 85 $\pm$24 & 1.10$\pm$0.21 & 10 & 0.70798  & 200$\pm$39 & 0.95$\pm$0.12 \\
           \noalign{\smallskip}
           4 & 0.60312 & 100$\pm$13 & 1.11$\pm$0.07 & 11 & 0.73368  & 127$\pm$31 & 1.00$\pm$0.17 \\
           \noalign{\smallskip}
           5 & 0.62500 & 133$\pm$19 & 1.00$\pm$0.07 & 12 &  0.74514 & 110$\pm$29 & 1.00$\pm$0.18 \\
           \noalign{\smallskip}
           6 & 0.63645 & 143$\pm$24 & 0.61$\pm$0.06 & 13 & 0.75660 & 108$\pm$32 & 0.91$\pm$0.19 \\
           \noalign{\smallskip}
           7 & 0.66215 & 170$\pm$25 & 0.95$\pm$0.08 &  14  & 0.76354  & 102$\pm$31 & 0.91$\pm$0.20 \\
           \hline
         \end{tabular}
         \begin{tablenotes}
         \item \emph{Notes}. The angular sizes are shown in Fig.~\ref{fig:3} (right panel). Cols.~2 and 6 -- time on October 5, 2023 in fractions of a day (UTC); Cols. 3 and  7 –- modeled flux density; Cols. 4 and 8 –- fitted angular size (FWHM of the circular Gaussian brightness distribution model component). 
         \end{tablenotes}
 %        \end{threeparttable}
   \end{table*} 

%% For this sample we use BibTeX plus aasjournals.bst to generate the
%% the bibliography. The sample631.bib file was populated from ADS. To
%% get the citations to show in the compiled file do the following:
%%
%% pdflatex sample631.tex
%% bibtext sample631
%% pdflatex sample631.tex
%% pdflatex sample631.tex

\bibliography{sample631}{}
\bibliographystyle{aasjournal}

%% This command is needed to show the entire author+affiliation list when
%% the collaboration and author truncation commands are used.  It has to
%% go at the end of the manuscript.
%\allauthors

%% Include this line if you are using the \added, \replaced, \deleted
%% commands to see a summary list of all changes at the end of the article.
%\listofchanges

\end{document}